\documentclass[10pt, conference, letterpaper]{IEEEtran}
\usepackage{booktabs}

\usepackage{flushend}

\ifCLASSINFOpdf
\else
  \usepackage[dvips]{graphicx}
  \graphicspath{{../eps/}}
  \DeclareGraphicsExtensions{.eps}
\fi

\ifCLASSOPTIONcompsoc
\usepackage[tight,normalsize,sf,SF]{subfigure}
\else
\usepackage[tight,footnotesize]{subfigure}
\fi
\usepackage[cmex10]{amsmath}
\usepackage{amssymb}

\usepackage{algorithmic}

\usepackage{paralist}
\usepackage{algorithm}
\usepackage{url}
\usepackage{multirow}

\makeatletter

\newcommand{\Rmnum}[1]{\expandafter\@slowromancap\romannumeral #1@}
\makeatother

\begin{document}
%
\title{Tars: Timeliness-aware Adaptive Replica Selection for Key-Value Stores}

\author{
\IEEEauthorblockN{Wanchun Jiang, Liyuan Fang, Haiming Xie, Xiangqian Zhou, Jianxin Wang}
\IEEEauthorblockA{
School of Information Science and Engineering, Central South University, Changsha, Hunan, China 410083\\
Email: jiangwc@csu.edu.cn
}
}
\maketitle

\begin{abstract}
In current large-scale distributed key-value stores, a single end-user request may lead to key-value access across tens or hundreds of servers. The tail latency of these key-value accesses is crucial to the user experience and greatly impacts the revenue. To cut the tail latency, it is crucial for clients to choose the fastest replica server as much as possible for the service of each key-value access. Aware of the challenges on the time varying performance across servers and the herd behaviors, an adaptive replica selection scheme C3 is proposed recently. In C3, feedback from individual servers is brought into replica ranking to reflect the time-varying performance of servers, and the distributed rate control and backpressure mechanism is invented. Despite of C3's good performance, we reveal the timeliness issue of C3, which has large impacts on both the replica ranking and the rate control, and propose the Tars (timeliness-aware adaptive replica selection) scheme. Following the same framework as C3, Tars improves the replica ranking by taking the timeliness of the feedback information into consideration, as well as revises the rate control of C3. Simulation results confirm that Tars outperforms C3.
\end{abstract}


%
\IEEEpeerreviewmaketitle

\section{Introduction}
In current large-scale distributed key-value store system, data is partitioned into small pieces, replicated and distributed across servers for parallel access and scalability. Consequently, a single end-user request may need key-value access from tens or hundreds of servers \cite{Dynamo, bing, facebook}. The tail latency of these key-value accesses decides the response time of the end-user request, which is directly associated with the user experience and the revenue \cite{latency, revenue}.
Nevertheless, because the performance of servers is time-varying \cite{AtScale,time-varying}, the tail latency is hard to be guaranteed, and may become long beyond expectation in certain condition.
Recent study shows that the $99^{th}$ percentile latency can be one order of magnitude larger than the median latency \cite{AtScale}, indicating  that there is a large space to cut the tail latency of key-value accesses.
To cut the tail latency, the replica selection scheme, which choose the fastest replica server for each key-value access as much as possible at clients, is crucial \cite{C3}.
Many other methods, including duplicate or reissue requests \cite{bing, AtScale, CosTLO, redundancy} for small tail latency, can also benefit from a good replica selection scheme.

However, the replica selection schemes of current classic key-value stores are very simple for efficiency.
For example, the OpenStack Swift just randomly reads from a server and retries in case of failures. HBase relies on HDFS, which chooses the physically closest replica server \cite{HDFS}. Riak uses an external load balancer such as Nginx \cite{Riak}, which employs the Least-Outstanding Requests (LOR) strategy. According to the LOR strategy, the client chooses the server to which it has send the least number of outstanding requests. MongoDB selects the replica server with smallest network latency \cite{mongodb}. Cassandra employs the dynamic snitching strategy, which considers the history of read latencies and I/O load \cite{Cassandra}.
Obviously, all these methods never take the time-varying performance of servers into consideration. Hence, they are hard to ensure the choice of the fastest replica server.

In spite of the time-varying performance of servers, the design of replica selection scheme still faces the following challenges. First, as all the clients independently choose the fastest server, they may concurrently access the fastest server, leading to great server performance degradation.
The same behavior will subsequently be repeated on a different fast server.
Therefore, this kind of herd behavior should be avoided by the replica selection algorithm.
Second, the replica selection scheme should be simple enough in the respect of both computation and coordination.
Aware of these challenges, an adaptive replica selection scheme C3 is proposed recently \cite{C3}.
C3 piggybacks the queue-size of waiting keys and the service time from the servers to guide the replica ranking at clients, and introduce both the Cubic rate control algorithm \cite{Cubic} and backpressure mechanism to adapt the sending rate of keys at the clients to the observed receipt capacity of servers.
In this way, C3 can adapt to the time-varying service rate across servers and avoid the herd behavior \cite{C3}.
The great benefit of the innovations on introducing the feedback, and the rate control and backpressure mechinism into the replica selection scheme is confirmed by both the experiments with Amazon EC2 and the at-scale simulations.

In this paper, we reveal the timeliness issue of C3, which has large impacts on both the replica ranking and the rate control. First, in the replica ranking of C3, when the network delay is ignored, the server with minimal $Q_s/\mu_s$ is the best candidate to cut the tail latency, where $Q_s$ denotes the queue-size of waiting keys at server and $\mu_s$ stands for the service rate of that server.
But our reproduced simulation shows the estimation accuracy of the $Q_s$ is poor in C3, especially when a concurrency compensation term $n*OS_s$ takes effect.
Detailed analysis reveals it is the poor timeliness of the feedback information that leads to the poor estimation accuracy, and the term $n*OS_s$ can not properly reflects the degree of concurrency.

Second, due to the timeliness of feedback information, congestion control algorithms for large delay are expected in key-value store. This may be why the Cubic rate control algorithm is utilized by C3. But the goal of rate control in C3 is to adapt the sending rate of keys to a server $s$, $sRate_s$, to the reception rate of returned values, $rRate_s$, from the server $s$. This is different from that of CUBIC, which adapts the sending rates of all clients to the total service capacity of server. Obviously, as the load of a server $s$ is decided by all these clients instead of a single one, $rRate_s$ can't reflect the total service capacity of server $s$.
Therefore, the goal of rate control in C3 should be revised.

Motivated by these observations, we propose the timeliness-aware replica selection (Tars) scheme, improving both the replica ranking and the rate control of C3 in this paper.
Tars follows the same framework as C3, and accordingly is simple enough for implementation. Different from C3, Tars piggybacks the incoming rate of keys $\lambda_s$ and the service rate $\mu_s$ from servers, and takes the timeliness of feedback information into consideration.
In replica ranking, Tars develops a scoring method without feedback information, when the timeliness of the feedback information is poor. When the feedback information is fresh, Tars estimates the queue-size more accurately with the help of feedback information $\lambda_s$ and $\mu_s$.
Moreover, Tars revises the goal of the rate control in C3,
making it consistent with the goal of the congestion control algorithms for Internet \cite{TCP, Cubic}.
Although the timeliness issue is not totally addressed, Tars outperforms C3 with these improvement, as confirmed by the simulations based on the open source code of C3. 

In sum, we make the following contributions in this paper:
\begin{asparaitem}
\item We reveal the timeliness issue of the framework developed by C3, and the drawbacks of C3 on replica ranking and rate control.
\item To address these issues, we propose the Tars scheme, which considers the timeliness of feedback information in replica ranking and revises the goal of rate control. Simulation confirms the advantages of Tars over C3.
\end{asparaitem}

The rest of this paper is organized as follows: Section II introduces the background. And then the motivation behind this work is presented in Section III. Subsequently, Section IV describes the design of the Tars scheme and Section V evaluates Tars with simulations based on the open source code of C3. Finally, Section VI concludes this paper.

\section{Background}
\begin{figure}[!t]
\centering
\includegraphics[width=3.1in]{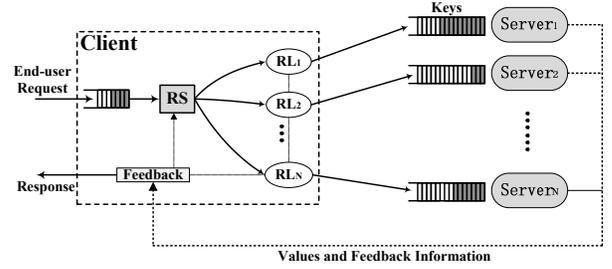}
\caption{Framework of C3. RS: Replica Selection. RL: Rate Limiter} \label{framework}
\end{figure}

In the key-value store, when a web sever receives an end-user request, it typically generates tens or hundreds of keys, and needs to access the corresponding values from different servers. The web server is also the client in the following key-value store, as shown in \figurename\ref{framework}.
For each key, the corresponding value is typically replicated and distributed across different servers.
When there is a key to send, client can find the corresponding replica servers via consistent hashing, and select a   replica server to send the key for the key-value access. Obviously, to cut the tail latency of key-value accesses, the fastest server is expected in the replica selection of each key at client.
On the other hand, a server can receive keys from different clients, and its service rates for keys are time-varying. When the server is busy, the newcome keys will be put into the waiting queue.
After a key is served, the corresponding value is returned to the client.

It is hard to ensure the choice of the fastest server for every key such that the corresponding value is returned as soon as possible. One reason is that the service time of keys are time varying, as the performance of server is influenced by many factors \cite{AtScale,time-varying}.
The other reason is that the size of the waiting keys at server is unknown, due to the large degree of concurrency in key-value access.
In other word, to know which server is the fastest, we not only need to obtain the network latency, but also have to capture the waiting time and the service time of keys at server.
Furthermore, the herd behavior, where the fast servers are preferred by most of the clients and get great performance degradation due to accompanying concurrent access, should be avoided.

Aware of these challenges, C3 suggests the server to monitor the queue-size of the waiting keys and its service time, and piggyback these information to client when the value is returned, as shown in \figurename\ref{framework}.
The feedback information is utilized for both the replica ranking and the rate control in C3.
Briefly, on the reception of a returned value, the client reads the feedback information and adjust the RL based on it via rate control algorithm.
When there is a key to be send, the client computes scores of each replica, ranks the replicas based on the scores via the RS scheduler, and then sequently inquires the states of RLs corresponding to these replicas.
If the current sending rate is within a RL, the corresponding replica is chosen to sent the key and the inquiry stops.
Or else, the following RL corresponding to a higher score replica is inquired.
If the current sending rate is not within all the RLs, the backpressure mechanism is triggered and the key is put into the backlog queue until there is at least one server within the RL again.

In the replica ranking of C3, the replica server with the smallest expected waiting time $\bar{q_s}*\bar{T_s}$ is preferred, where $\bar{T_s}$ is the EWMAs of the feedback service time $T_s$ of a key and $\bar{q_s}$ is the queue-size estimation of the waiting keys. $\bar{q_s}$ is defined as follows.
\begin{equation}
\bar{q_s}\triangleq1+q_s+n*os_s
\label{C3-est}
\end{equation}
Here $q_s$ is the EWMAs of the feedback queue-size $Q^f_s$, $n$ is the number of client, and $os_s$ is the number of outstanding keys whose values are not yet to be returned.
In equation (\ref{C3-est}), the term $n*os_s$ is considered as the concurrency compensation \cite{C3}.
The specifical scoring function used for replica ranking of C3 is as follows.
\begin{equation}
\Psi_s=\bar{R_s}-\bar{T_s}+\bar{q_s}^3*\bar{T_s}
 \label{rank}
\end{equation}
where $\bar{R_s}$ is the Exponentially Weighted Moving Averages (EWMAs) of the response times witnessed by client, and thus the $\bar{R_s}-\bar{T_s}$ is the considered as the delay.
Moreover, the term $\bar{q_s}^3*\bar{T_s}$ is the replacement of $\bar{q_s}*\bar{T_s}$ in order to penalizing long queues in Eq. (\ref{rank}), and the mechanism is named as Cubic replica selection in C3.
The replica server with the smallest $\Psi_s$ is selected by the RS scheduler, when a key is going to be sent.

The rate control and backpressure mechanism is as follows. As shown in \figurename\ref{framework}, a client maintains a Rate Limiter (RL) for each server to limit the number of keys sent to the server within a specified time interval $\delta$, named $sRate_s$. The key will not be sent to a server when the corresponding rate is limited.
If the rates of all the replica servers of a key are limited, the key will be put into a backlog queue until the rate limitation of a replica server is released. The detailed rate control algorithm is borrowed from CUBIC \cite{Cubic}. Let $rRate_s$ be the number of values received from a server in a $\delta$ interval. $sRate_s$ is increased according to the following cubic function when $sRate_s<rRate_s$.
\begin{equation}
sRate\rightarrow \gamma*(\Delta T-\sqrt[3]{\frac{\beta*R_0}{\gamma}})^3 + R_0
\end{equation}
wherein $R_0$ is the recorded $sRate_s$ before the previous rate-decrease, $\Delta T$ is the elapsed time since the previous rate-decrease event, and $\gamma$ is constant coefficient.
When $sRate_s>rRate_s$, and a $2*\delta$ hysteresis period after the rate increase, $sRate_s$ is decreased to $\beta*sRate_s$, where $\beta$ is a positive constant smaller than 1.
The hysteresis period $2*\delta$ is enforced for the measurement of $rRate_s$ after a rate increase.
The rate adjustment is done on the receipt of each returned value, aiming to adapt the $sRate_s$ to the $rRate_s$, but the rate adjustment result will take effect when there are keys to be sent.

With the cooperation of the replica ranking method and the rate control and backpressure mechanism, C3 can adapt to the time-varying service time across servers and avoid the herd behavior, and accordingly achieve high throughput and low tail latency, as confirmed by experiments and simulations in \cite{C3}.

\section{Motivation}
Although C3 has great innovation on bringing feedback into the replica ranking and developing the rate control and backpressure mechanism, the detailed replica ranking method and rate control algorithm can be further improved. Specifically, we find the timeliness issue of C3, and the drawbacks of C3 on the estimation of queue-size in the replica ranking and the goal of rate control.
For the convenience of reading, we summarize the key notations used in this paper in Table \ref{notations}.

\begin{table}
\caption{Definitions of key notations}
\centering
\begin{tabular}{c|c}
\toprule[2pt] Notation      & Definition\\
\midrule[1pt] $Q_s$          & The real queue-size of waiting keys at server $s$\\
\hline       $Q^f_s$         & The feedback queue-size from server $s$\\
\hline       $q_s$           & The EWMAs of $Q^f_s$ computed in C3\\
\hline       $\bar{q}_s$      & The estimation of queue-size of server $s$ at client\\
\hline \multirow{2}*{$\tau_w$} & The interval from the reception of feedback information \\
                               &  and the use of it for current replica selection\\
\hline  \multirow{2}*{$f_s$}   & The number of times that the replica server s  \\
                               & is not selected during the time interval $\tau_w$ \\
\hline         {$\tau^s_w$}    & The time of the corresponding key staying at server $s$\\           
\hline         {$T_s$}         & The feedback service time of the corresponding key at server $s$\\ 
\hline         {$\mu_s$}         & The feedback service rate of keys measured at server $s$\\ 
\bottomrule[2pt]
\end{tabular}
\label{notations}
\end{table}

\begin{figure}[!t]
\centering
\includegraphics[width=2.4in]{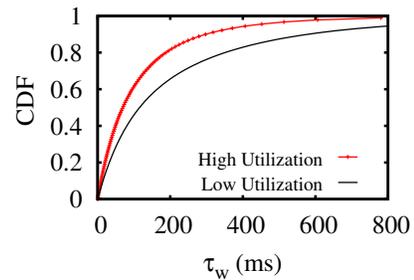}
\caption{CDF of the $\tau_w$ in C3.} \label{waittime}
\end{figure}

\subsection{Timeliness of Feedback}
The feedback information plays a key role in above framework of replica selection developed by C3. However, we find the timeliness of the feedback information may be poor frequently. More specifically, the feedback information would be delayed for a propagation time $\tau_d$ before it arrives at the client, and there is also a time interval $\tau_w$ during the reception of feedback information and the utilization of this feedback information for current replica selection.
In fact, we find the value of $\tau_w$ can vary in a large range due to the following reasons.
First, after a client receives feedback information from a server, it may not send keys to this server for a long time, either because this server doesn't belong to the replica group of the following keys sent by this client, or because this server is not selected due to its poor performance.
In this condition, the feedback information can't be renewed timely.
Second, even if the client sends key to a server after receiving the feedback information from it, feedback information will be renewed when the value of this key is returned. Obviously in this case, the value of $\tau_w$ is larger than the latency of this key-value accesses. As the $99^{th}$ percentile latency of key-value accesses can be one order of magnitude larger than the median latency \cite{AtScale}, the value of $\tau_w$ could also change in a very large range.
To exhibit the timeliness of the feedback information, we reproduce simulations in C3 (see part A of section V for detailed simulation configuration), and collect the values of $\tau_w$ before the sending of each key. \footnote{600000 values are collected. After the CDF is computed, we present only 5\% of data to reduce the size of figure, without changing the sharp of curves} The cumulative distribution function of $\tau_w$ is shown in \figurename\ref{waittime}.
Consisted with above insights, the $\tau_w$ has a very large probability to become as large as hundreds of milliseconds, especially when the server utilization is low, while the network latency $\tau_d$ is only in the order of several milliseconds.
Therefore, the timeliness of feedback information in above framework is poor.
This maybe also the reason why the replica selection algorithms in current classic key-value stores all don't heavily rely on feedback information.
Subsequently, we will focus on the impacts of the timeliness of feedback information on the replica ranking and rate control of C3.

\begin{figure}[!t]
\centering
\includegraphics[width=2.5in, angle=-90]{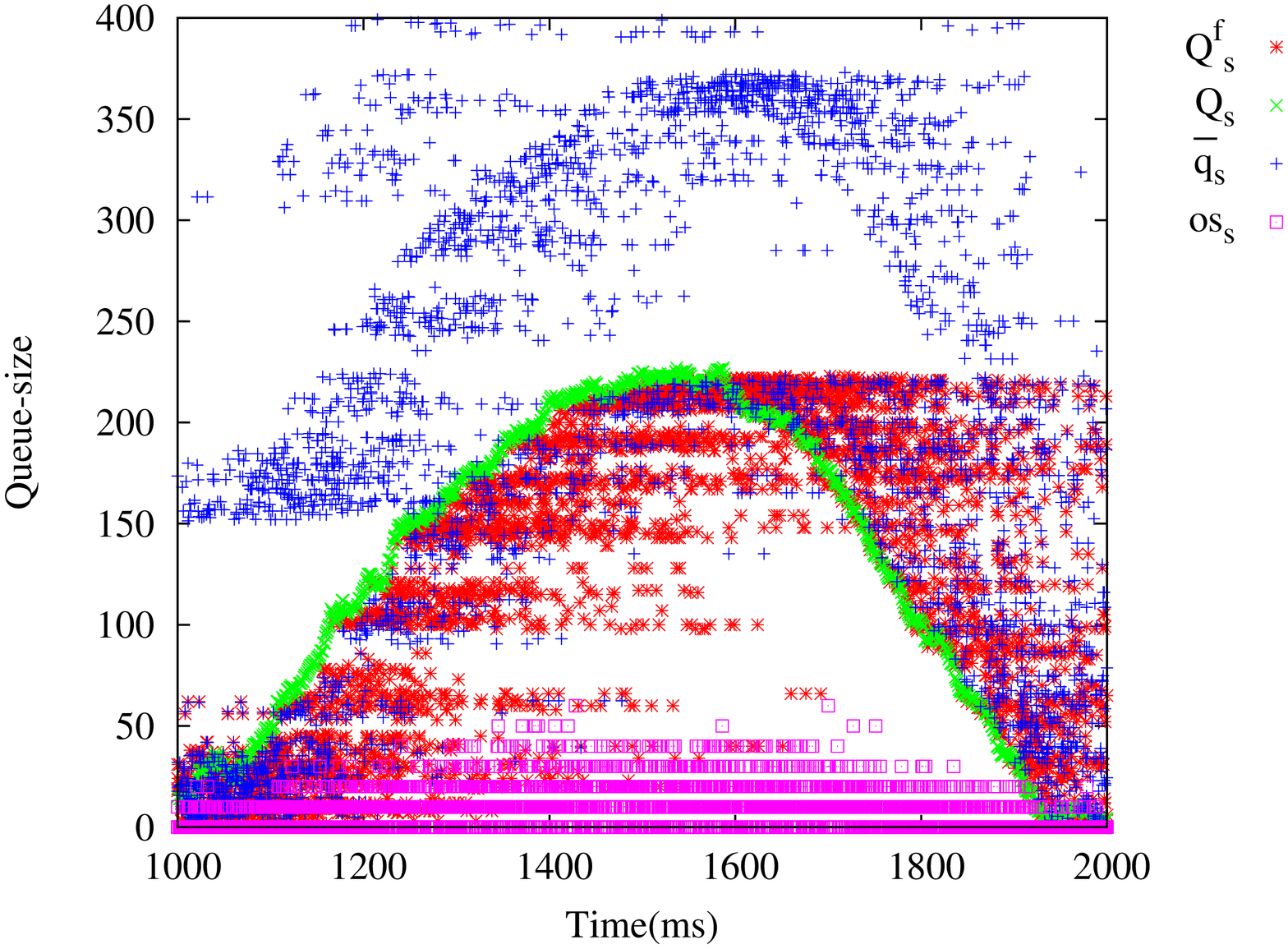}
\caption{Queue-size and its estimation in C3. Values exceed 400 are not shown and $os_s$ is multiplied by 10 for the convenience of observation.} \label{queue}
\end{figure}

\subsection{Replica Ranking}
\begin{figure}[!t]
\centering
\includegraphics[width=2.5in, angle=-90]{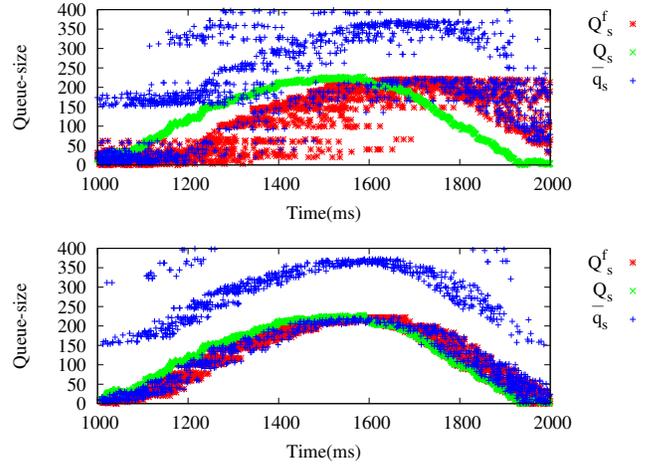}
\caption{Queue-size and its estimation in C3 with different values of $\tau_w$. \quad (Up: $\tau_w> 100$  ms, Down: $\tau_w\leq 100$  ms)}
\label{queuenew}
\end{figure}

Due to the poor timeliness of feedback information, the estimation accuracy of the queue-size of the waiting keys and the service time, both of which are crucial for the replica ranking of in C3, is poor.
Specifically, as shown in \figurename\ref{queue}, we randomly choose a server $s$ to show its queue-size of waiting keys $Q_s$ at each time when the scoring is executed at clients, as well as all of the feedback $Q^f_s$ received from server $s$ at clients, the $os_s$ and its estimation $\bar{q_s}$ on the queue-size of server $s$ in a random simulation time interval.
There is a large difference among the piggybacked queue-size $Q^f_s$, the estimation $\bar{q_s}$ and the real queue-size $Q_s$.
The large degree of concurrency in key-value access is considered as one of the main reason for this phenomenon, and accordingly the term $n*os_s$ is utilized as the concurrency compensation in the computation of the estimated queue-size $\bar{q}_s$, as represented in C3 \cite{C3}.
However, the term $n*os_s$ has not helped to improve the estimation accuracy of the queue-size, as illustrated in \figurename\ref{queue}.
In fact, dividing the data of \figurename\ref{queue} into two sub figures with threshold $100$ ms on $\tau_w$,
we show \emph{it is the poor timeliness of feedback information that leads to the poor estimation accuracy of the queue-size}.
Specifically, as shown in \figurename\ref{queuenew}, the difference among the real queue-size $Q_s$, its estimation $\bar{q}_s$ and the feedback queue-size $Q^f_s$ is small when $\tau_w\leq 100 ms$, excepting the condition that $os_s$ is nonzero. When the value of $\tau_w$ becomes in the order of hundreds of milliseconds, the real queue-size $Q_s$ can change greatly during such a large time interval, and thus cann't be estimated based on the old feedback information.
Therefore, when $\tau_w$ is large, the replica selection methods independent of feedback information are needed.
Similarly, the timeliness of the feedback service rate of servers may also becomes poor frequently.

Furthermore, when $\tau_w$ is small, the queue-size may also change a lot due to the large degree of concurrency in key-value access.
The term $n*os_s$ can not properly represent the degree of concurrency, as
the degree of concurrency will be constrained by the rate control algorithm. Hence, it is not reasonable to assign the weight $n$ to $os_s$.
In fact, we find the term $n*os_s$ is helpful in simulation, not because it compensates the impact of concurrency and makes the queue-size estimation better, but because that the corresponding server should not be chosen before the outstanding keys are served and the feedback information is piggybacked and renewed.
To improve the queue-size estimation in this condition, we suggest to piggyback some better variables as the feedback information except for $Q^f_s$ and $T_s$.

\subsection{Rate Control}
We also find that the timeliness of feedback information has great impact on the rate control of C3.
Although the rate adjustment is executed immediately after the feedback information is received from a server, this rate adjustment doesn't make sense if the client doesn't send any key to this server for a relatively long time interval.
This is much different from the congestion control of Internet, which assumes there are always data to send.
Even if the client sends keys to this server right after the rate adjustment, i.e., the rate adjustment results take effect on time, the congestion control algorithms faces a forward time delay $\tau_w$, which denotes the timeliness of feedback information.
Note that the value of $\tau_w$ can change in a very large range, i.e., from several milliseconds to hundreds of milliseconds.
This kind of delay would has great impacts on the stability of rate control algorithms.
This may be why C3 adopts the CUBIC algorithm, which is designed for networks with large bandwidth delay product.

Although the distributed rate control is inspired by congestion control of Internet, the goal of rate control in C3 is not suitable for the key-value stores.
Specifically, in C3, the $rRate_s$ is used to represent the perceived performance of a server $s$, and the goal is to adapt the $sRate_s$ to the $rRate_s$ at clients.
The benefit is that no feedback is needed, because the $rRate_s$ can be independently measured at client.
However, the $rRate_s$ can only reflect the service capacity of server $s$ allocated to this client, while the service capacity of server $s$ is competed by many clients, as it accepts keys from many different clients.
The $rRate_s$ may not be able to reflect the total service capacity of servers.
This is different from the CUBIC algorithm for Internet, which adapt the sending rates of all clients to the total service capacity of servers.
In CUBIC, the total service capacity of servers is reflected by whether the buffer overflows. 
Hence, the goal of rate control in C3 should be revised.

%

In a word, we reveal the timeliness issue of the replica selection framework developed by C3, and the drawbacks of C3 on the replica ranking and the goal of rate control.

\section{Design of Tars}
Motivated by above insights, we design the Tars scheme, which follows the same framework as C3, but improves the replica ranking and rate control methods. The specific improvements are as follows.

\subsection{Timeliness-aware Replica Ranking}
The procedure of replica ranking of Tars is the same as that of C3, illustrated in \figurename\ref{framework}. But Tars has  different feedback information and scoring methods.

\noindent\textbf{Feedback Information}
In contrast to C3, where the queue length $Q^f_s$ and the service time $T_s$ are piggybacked, Tars utilizes the following feedback information: the queue length $Q^f_s$, the incoming rate of keys $\lambda_s$, the service rate $\mu_s$ and the time of the key staying at the server $\tau^s_w$. Obviously, $\tau^s_w$ is the sum of the service time $T_s$ and the queuing time of the key at the server. Note that $\mu_s$ is different from $T^{-1}_s$ when the server can concurrently process several keys, as discussed in part A of section V.
The $T_s$ is never used again, and replaced by $\tau^s_w$ and $\mu_s$ in Tars

\noindent\textbf{Timeliness of Feedback}
As discussed in section III, the timeliness of feedback is represented by $\tau=\tau_d+\tau_w$.
Obviously, the duplex network delay can be computed by $\tau_d=R_s-\tau^s_w$, where $R_s$ is the response time witnessed by client, but without EWMAs, and $\tau^s_w$ is involved in the feedback information.
Moreover, the initialization of time interval $\tau_w$ is the time when a client receives a returned value and the feedback information is extracted. The end of time interval $\tau_w$ is the current time when a new key is going to be sent based on the replica ranking utilizing this feedback information.
Because $\tau_d$ is only in the order of several milliseconds and $\tau_w$ can become as large as hundreds milliseconds, Tars mainly uses $\tau_w$ to represent the timeliness of feedback information.
When the timeliness of feedback information is poor, Tars develops a scoring method independent of feedback information.
Conversely, Tars is inclined to estimate the queue-size of waiting keys and the service rate more accurately, and employs a scoring method similar to C3.
Referring to the dynamic snitch mechanism of Cassandra, the $100$ ms is chosen to be the boundary of utilizing different scoring methods.

\noindent\textbf{Queue-size Estimation}
When $\tau_w\leq100$ ms, the scoring method based on queue-size estimation method is adopted in Tars.
Specifically, Tars assumes both $\lambda_s$ and $\mu_s$  changes a little during time interval $\tau_d$ in this condition, and then computes the queue-size with the following approximation.
\begin{equation}
Q^f_s+(\lambda_s-\mu_s)*\tau_d\approx Q_s
\label{queue-est-principle}
\end{equation}
where $Q_s$ is the real queue-size of waiting keys at server $s$.
Note that $\tau_w$ is not involved in (\ref{queue-est-principle}), because the rates $\lambda_s$ and $\mu_s$  may change in a relatively large time interval due to the large degree of concurrency.
Obviously, equation (\ref{queue-est-principle}) is also hard to accurately estimate the real queue-size.
But comparing equation (\ref{queue-est-principle}) to equation $(\ref{C3-est})$, where the queue-size estimation of C3 becomes $1+q_s$ without taken the term $os_s$ into consideration, the term $(\lambda_s-\mu_s)*\tau_d$ can be considered as the concurrency compensation and equation (\ref{queue-est-principle}) can be a better queue-size estimation method than $1+q_s$.
In addition, similar to C3, the term $n*os_s$ is also added in the queue-size estimation of Tars, based on the intuitive viewpoint ``the replica server is not preferred when there are already keys sent to this server but without returned value'', instead of being considered as the concurrency compensation.
In total, the queue-size estimation about server $s$ in Tars is.
\begin{equation}
\bar{q_s}=Q^f_s+(R_s-\tau^s_w)(\lambda_s-\mu_s)  + n*os_s
\label{queue-est}
\end{equation}
Note that different from C3, all variables are utilized directly without EWMAs in equation (\ref{queue-est}), excepting $\lambda_s$ and $\mu_s$, because the EWMAs brings in some more staler feedback information.

\noindent\textbf{Scoring with Feedback}
When $\tau_w\leq100$ ms, the replica ranking of Tars uses the following scores based on the queue-size estimation (\ref{queue-est}).
\begin{equation}
\Psi_s=R_s-\tau^s_w+\bar{q_s}^3/\mu_s
\label{rank-r3}
\end{equation}
Compared with equation (\ref{rank}) and (\ref{rank-r3}), we can find that the difference between the scoring methods of  C3 and that of Tars are triple.
\begin{asparaitem}
\item First, the term $T_s$ is replaced by $\tau^s_w$, i.e., the waiting time of the key at server is not considered as the access latency in Tars, because $\bar{q_s}^3/\mu_s$ stands for it.
\item Second, the queue-size estimation methods are different from each other, as Tars takes the timeliness of feedback information into consideration.
\item Third, as the server can concurrently process several keys, the service rate is measured independently in Tars, instead of using the reciprocal of the service time $T^{-1}_s$.
\end{asparaitem}

\begin{algorithm}[!t]
  \caption{Scoring on server $s$ (When there is a key to send)}
  \label{score}
  \begin{algorithmic}[1]
  \STATE Compute $\tau_w$
  \IF{$\tau_w>100$  ms}
    \IF {$os_s == 0$ and $f_s==0$}
        \STATE $\bar{q_s} \gets 0$   \quad // no key to send for a long time
    \ELSIF {$os_s == 0$ and $f_s>6$}
        \STATE $\bar{q_s} \gets 0$   \quad // replica is always not selected, try it
    \ELSE
        \STATE $\bar{q_s} \gets 1+q_s+os_s*n$ 
    \ENDIF
  \ELSE
    \STATE $\tau_d \gets R_s-\tau^s_w$
    \STATE $\bar{q_s} \gets Q^f_s + (\lambda_s-\mu_s)*\tau_d + os_s*n$
  \ENDIF
  \STATE $Score \gets R_s-\tau^s_w+\bar{q_s}^3/\mu_s$
  \end{algorithmic}
\end{algorithm}

\noindent\textbf{Scoring without Feedback}
When $\tau_w > 100$ ms, the feedback information become useless with the time elapse, and Tars develops the following scoring method without feedback for this condition.
Obviously, $\tau_w > 100$ ms indicates that the client has not sent any keys to server $s$ for a long time.
Let $f_s$ be the number of times that the replica server $s$ is not selected during the time interval $\tau_w$ recorded by client.
When $os_s=0$ and $f_s=0$, there is no key to be sent to the group of replica servers, where server $s$ belongs to, for a long time $\tau_w$ due to the traffic pattern. The client tends to send current key to server $s$ in this condition.
When $os_s=0$ and $f_s>6$, the replica server $s$ has not been selected for many times in a long time $\tau_w$, we send a key to this replica server to try whether this performance of this replica server has recovered.
Or else, Tars uses the same queue-size estimation method (\ref{C3-est}) as C3, because we don't have any more information.

Putting everything together, we can obtain the detailed scoring method of Tars utilized in replica ranking before sending keys, as shown in \textbf{Algorithm \ref{score}}.

\subsection{Rate Control}

As discussed in section III, the goal of the rate control in Tars is changed to adapt the sending rates of clients to the service rate of servers.
It means the sending rate of a client is decreased or increased based on whether the server is saturated or not in Tars.

\noindent\textbf{Rate Decrease}
The saturation state of servers, or the service capacity of servers can be reflected by whether the queue-size $Q^f_s$ is larger than a predefined value, i.e., whether there is a `` buffer overflow ''.
Different from $C3$, where the $sRate_s$ is decreased when $sRate_s > rRate_s$, Tars decreases the $sRate_s$ when the queue-size $Q^f_s$ exceeds a predefined value $B=5$, corresponding to the packet drops resulted by buffer overflows in the congestion control of the Internet.
The same as CUBIC and C3, the multiplicative rate decrease method is employed here, i.e., $sRate_s\leftarrow \beta*sRate_s$, 
where $\beta$ is a fixed coefficient smaller than 1.

\begin{algorithm}[!t]
  \caption{Rate Adjustment (on reception of returned value)}
  \label{rate-control}
  \begin{algorithmic}[1]
  \STATE record current time as the initialization of $\tau_w$
  \STATE initialization $f_s \gets 0$
  \STATE measure the response time $R_s$ to compute $\tau_d$
  \STATE update EWMA of $q_s$, $T_s$
  \IF { \textbf{($Q^f_s > 5$) } and ($now()-T_{inc} > 2*\delta$) } \label{r3-control}
    \STATE // ($sRate_s > rRate_s$) and ($now()-T_{inc} > 2*\delta$) \label{c3-control}
    \STATE $R_0 \gets sRate_s$ \textbf {when $\beta*sRate_s > 0.01$}
    \STATE $sRate_s \gets max(\beta*sRate_s, 0.01)$
    \STATE $T_{dec} \gets now()$
  \ELSIF {($sRate_s < rRate_s$)}
    \STATE $\Delta T \gets now()-T_{dec}$
    \STATE $T_{inc} \gets now()$
    \STATE $R \gets \gamma*(\Delta T-\sqrt[3]{\frac{\beta*R_0}{\gamma}})^3 + R_0$
    \STATE $sRate_s \gets min(sRate_s+s_{max}, R)$
  \ENDIF
  \end{algorithmic}
\end{algorithm}

\noindent\textbf{Rate Increase}
In contrast that the sending rate is increased periodically after the rate decrease in the Cubic congestion control of the Internet, Tars does not increase the $sRate_s$ whenever $sRate_s\geq rRate_s$.
This is because $rRate_s$ reflects the real sending rate of client to server $s$, and $sRate_s$ is the boundary of the rate limiter for server $s$. When $sRate_s\geq rRate_s$, all the keys can be sent without rate limiting, and thus it's meaningless to further increase the value of $sRate_s$ in this condition.
Therefore, $sRate_s$ is only increased when it's smaller than $rRate_s$ in Tars.


Putting above viewpoints together, we can obtain the detailed rate control algorithm of Tars, as shown in \textbf{Algorithm \ref{rate-control}}.
In fact, the rate control algorithm \ref{rate-control} is almost the same as that of C3. The major difference is that the judgement condition for rate decrease (step \ref{c3-control}) is replaced by the step \ref{r3-control}.
Another improvement made by Tars is in step 7, which ensures that the target value $R_0$ for rate increase never reaches the lower bound value of $sRate_s$.
In addition, step 1 and step 2 are newly added in Tars.


\subsection{Discussion}
Compared to C3, Tars utilizes the same framework and has similar replica ranking and rate control methods. Hence, Tars is also simple and implementable, can avoid the herd behavior, and is adaptive to the time-varying performance across servers, similar to C3.

In reality, because of the large degree of concurrency and the poor timeliness of feedback, it's hard to accurately estimate the queue-size of waiting keys of servers, especially when $\tau_w > 100$ ms.
Note that when $\bar{q_s}$ is always set as $0$, both C3 and Tars will degenerate to the replica selection scheme, where the server with the smallest network latency is chosen.
Obviously, there is larger probability to obtain a smaller estimation error when $\bar{q_s}$ is set the value of feedback queue-size $Q^f_s$, compared with letting $\bar{q_s}=0$.
Moreover, we believe the queue-size estimation equation (\ref{queue-est}) is better than equation (\ref{C3-est}), when $\tau_w\leq 100$ ms.

Excepting the goal of rate control, the rate control algorithm for key-value store also suffer from the timeliness issue as discussed in section III.
Therefore, there is chance to improve the rate control algorithm for key-value stores.
In this paper, we just revise the goal of rate control in C3, but leave the improvement on rate control algorithm as the further work.
Even with this small modification, the rate control of Tars becomes better than that of C3, as confirmed by simulation results in section V.

The distribution of $\tau_w$ is impacted by several factors. The most intuitive factors are as follows. First, the larger the workload, the larger probability that $\tau_w$ is of small values, as shown in \figurename{\ref{waittime}}.
In addition, the larger the number of clients, the smaller probability that $\tau_w$ is of small values, as the time interval for a client to receive feedback information becomes large.

\section{Evaluation}

\subsection{Implementation and Setup}
\noindent\textbf{Setup}
We implement Tars based on the open source code of C3 \cite{C3}. As in C3, the workload generators create keys at a set
of clients according to a Poisson arrival process to mimic arrival of user requests at web servers \cite{facebook}.
These keys are sent to a set of servers, each of which is chosen according to the replica selection algorithm at client from 3 replica servers.
The server maintain a FIFO queue for waiting keys, but can serve a tunable number (4 by default) of requests in parallel.
The service time of each key is drawn from an exponential distribution with a mean service time $T_s=4$ ms as in \cite{redundancy}.
The time-varying performance fluctuations of servers is simulated by a bimodal distribution \cite{server-model} as follows: each server sets its mean service rate either to $T^{-1}_s$ or to $D*T^{-1}_s$ with uniform probability every fluctuation interval $T$ ms, where D is a range parameter with default value 3.
The arrival rate of keys corresponds to 70\% (high utilization scenario, used by default) and 45\% (low utilization scenario) of the average service rate of the system.

\noindent\textbf{Service Rate}
Different from $C3$, we mainly modify the feedback information, the replica ranking and the rate control. Specifically, we revise the measurement method of service rate in C3.
In the code of C3, the service time of one key is returned directly and its reciprocals is considered as the service rate.
But each server serves keys in parallel to model the concurrent processing of multicore computer.
The macroscopical service rate of server is larger than the reciprocals of the service time of one key.
Therefore, to measure the service rate, we count the number of keys served during the service time of one key and piggyback it to the returned values for this key-value access.
Not that the service time may be very small such that there is no key served. In this condition, we take the number of keys served in two consecutive service time into consideration.
Similar method is used to measure the incoming rates $\lambda_s$ of keys at server.
Note that $\lambda_s$ and $\mu_s$ are always measured within the same time interval.

\noindent\textbf{EWMAs} In C3, the EWMAs of feedbacks are utilized to replace the original ones at clients. However, as the consecutive feedbacks are sent to different clients at a server, there may be a great difference between the old feedbacks and the fresh one. Hence, Tars utilize the feedbacks directly, excepting that the EWMAs of $\lambda_s$ and $\mu_s$ are computed at server before they are piggybacked.

\noindent\textbf{Configuration}
The configurations and metrics are the same as C3. 200 workload generators, 50 servers and 150 clients are used by default. The one-way network latency is 250 $\mu s$. The parameters of the Cubic function are $\beta=0.2$, $\gamma$ is set to be $0.000004$ such that the saddle region of Cubic function is $100$ ms, the unit of $\Delta T$ is ms and $s_{max}=10$.
The $99^{th}$ percentile latency is computed by taking the average of 5 repeated experiments, where different random seeds  are set and 600,000 keys are generated in each run.
Without declared explicitly, the high utilization scenario with $T=500$ ms is used.

\begin{figure}[!t]
\centering
\includegraphics[width=2.8in]{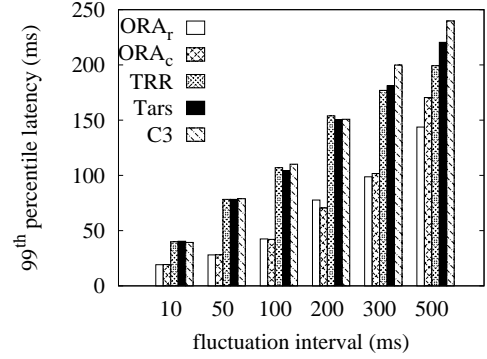}
\caption{Impacts of the time varying service rate.} \label{c150h}
\end{figure}

\noindent\textbf{Comparative} We mainly compare Tars to C3, as well as the following Oracle strategy. With the Oracle strategy, each client is assumed to has perfect knowledge of the instantaneous value $Q_s/\mu_s$ at each replica server.
Note that the Oracle strategy may be composed with rate control methods of C3 or Tars, named as $ORA_c$ and $ORA_r$ respectively.
For more detailed comparison, we also compose the timeliness-aware replica ranking of Tars and the rate control of C3 as one of the comparative, named TRR.

\begin{figure}[!t]
\centering
\includegraphics[width=2.8in]{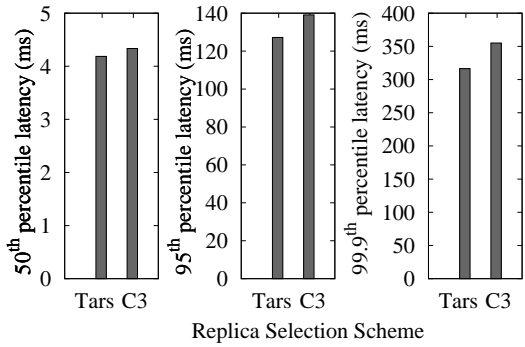}
\caption{Latencies of C3 and Tars.} \label{latency}
\end{figure}

\begin{figure}[!t]
\centering
\includegraphics[width=2.8in]{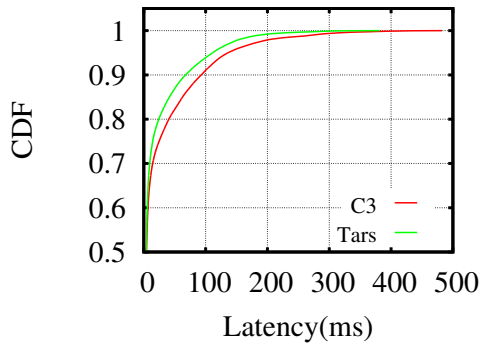}
\caption{CDF of the Latency of C3 and Tars.} \label{cdflatency}
\end{figure}

\subsection{Simulation Results}
In all the following simulation results, the $99^{th}$ percentile latencies of C3 is almost the same as that in the Fig.14 and Fig. 15 of $\cite{C3}$. This can serve as the evidence for the correctness of our implementation.

\noindent\textbf{Impacts of Time-varying Service Rate.}
As both C3 and Tars are designed to be adaptive with the time-varying performance across servers, we evaluates Tars with time varying service rate at first.
With the fluctuation interval $T$ of the average service time of servers changes from $500$ ms to $10$ ms, the $99^{th}$ percentile latencies are shown in \figurename\ref{c150h}.
With the same rate control and backpressure mechanism of C3 but different replica ranking methods, the $99^{th}$ percentile latencies of schemes satisfy $ORA_c \ll TRR < C3$.
It indicates that the tail latency can be cut greatly with perfect knowledge of the queue-size and the service time, but the queue-size estimation of both C3 and Tars are not very good, as discussed in part C of section IV. 
But the timeliness-aware replica ranking method of Tars is a little better than that of C3, as illustrated in \figurename\ref{c150h}.
Note that the difference among the $99^{th}$ percentile latencies becomes significant, when the time interval $T$ is a large value like $500$ ms, i.e., the average service time of servers is not changed frequently.
When $T=10$ ms, i.e., the average service time of servers changes frequently, the feedback information becomes stale very fast.
Correspondingly, the replica ranking based on feedback information becomes poor, and the rate control can't adapt to the rapid change of service capacity in both Tars and C3.
Therefore, the difference between Tars and C3 is small with $T=10$ ms.

\begin{figure}[!t]
\centering
\includegraphics[width=2.7in]{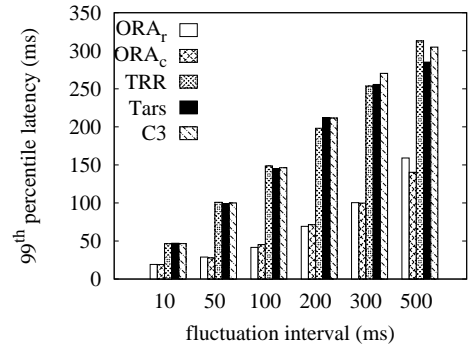}
\caption{Impacts of the number of clients. (n=300)} \label{c300h}
\end{figure}

\begin{figure}[!t]
\centering
\includegraphics[width=2.7in]{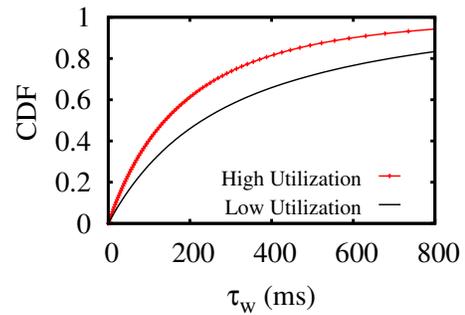}
\caption{CDF of $\tau_w$ with $n=300$ in C3.} \label{cdf300}
\end{figure}

\begin{figure*}[!t]
\centering
\subfigure[n=150]{
\includegraphics[width=2.7in]{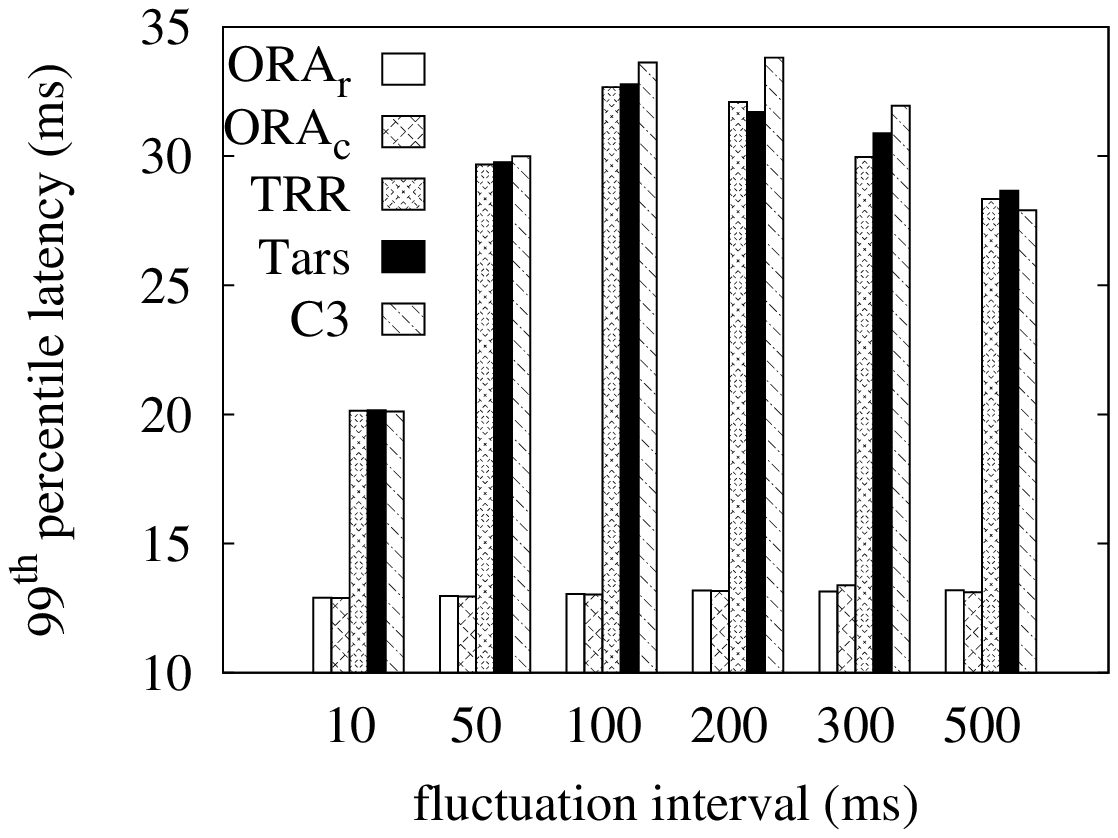}
}
\quad
\subfigure[n=300]{
\includegraphics[width=2.7in]{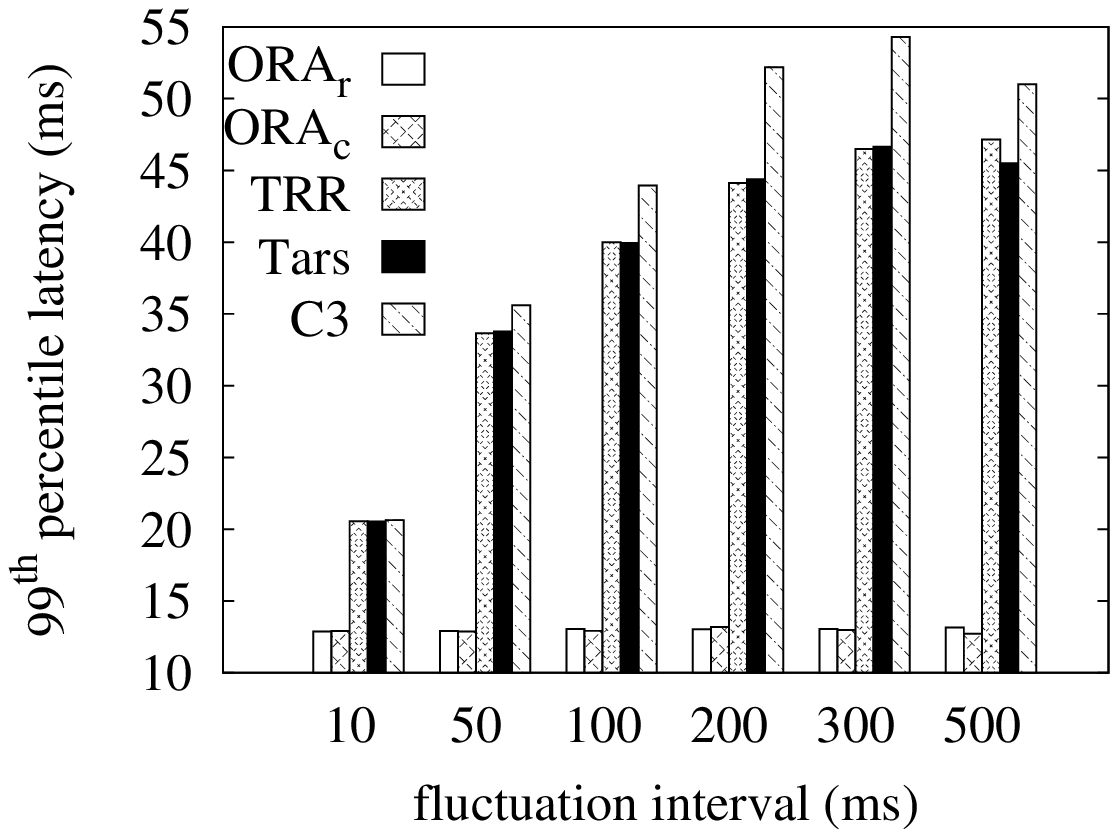}
\label{c300l}
}
\caption{Impacts of the server utilization. (utilization=0.45)}
\label{utilization}
\end{figure*}

Similarly, with the same replica ranking method but different goals of rate control, the $99^{th}$ percentile latency of schemes satisfy $ORA_r<ORA_c$ and $TRR<C3$. It indicates the rate control method of Tars is a little better than that of C3, with the revised goal of rate control.
Especially when $T=500$ ms, the rate control method of Tars is helpful when it cooperates with the $ORA$ strategy.

Finally, combining the timeliness-aware replica ranking and the revised goal of rate control, Tars always outperforms C3, as shown in \figurename\ref{c150h}.

\noindent\textbf{Latency}
To compare the performance of C3 and Tars in detail, we also illustrate the $50^{th}$ percentile latencies, the $95^{th}$ percentile latencies and the $99.9^{th}$ percentile latencies in \figurename\ref{latency}, when $T=500$ ms.
Under all of these metrics, Tars outperforms C3, and the advantage of Tars becomes the most significant with the metric $99.9^{th}$ percentile latency.
In fact, the CDF of the latencies of all key-value accesses can illustrate the advantage of Tars over C3 better, as shown in \figurename\ref{cdflatency}.

\begin{figure*}[!t]
\centering
\subfigure[n=150]{
\centering
\includegraphics[width=2.7in]{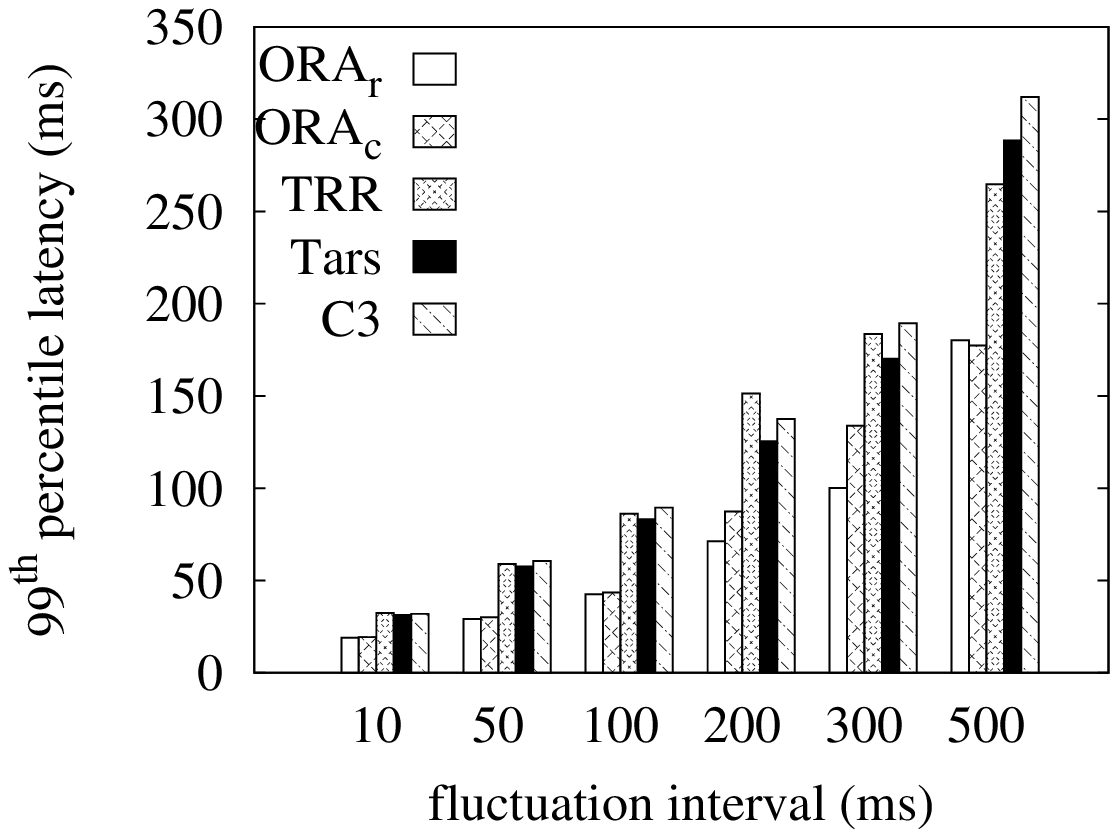}
}
\quad\quad\quad
\subfigure[n=300]{
\centering
\includegraphics[width=2.7in]{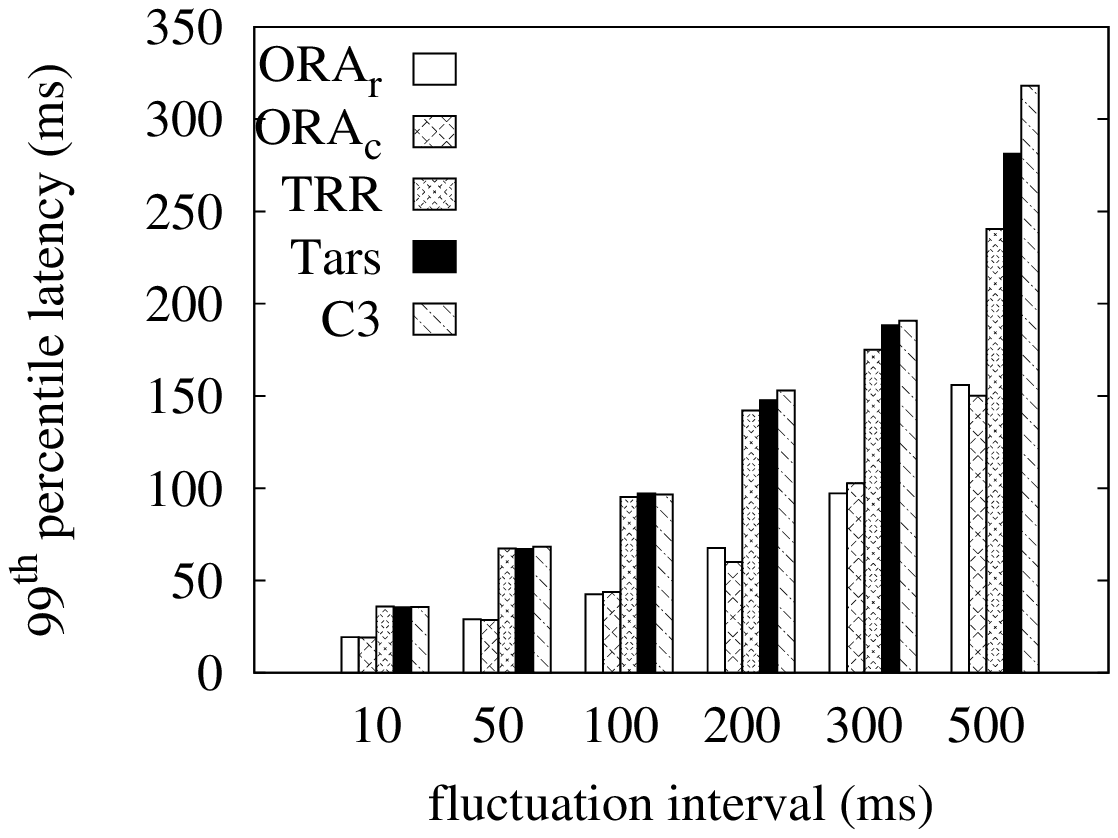}
}
\caption{Impacts of the skewed client demands: 20\% clients generate 80\% of the total demand.} \label{skew20}
\label{low-utilization}
\end{figure*}

\begin{figure*}[!t]
\centering
\subfigure[n=150]{
\centering
\includegraphics[width=2.7in]{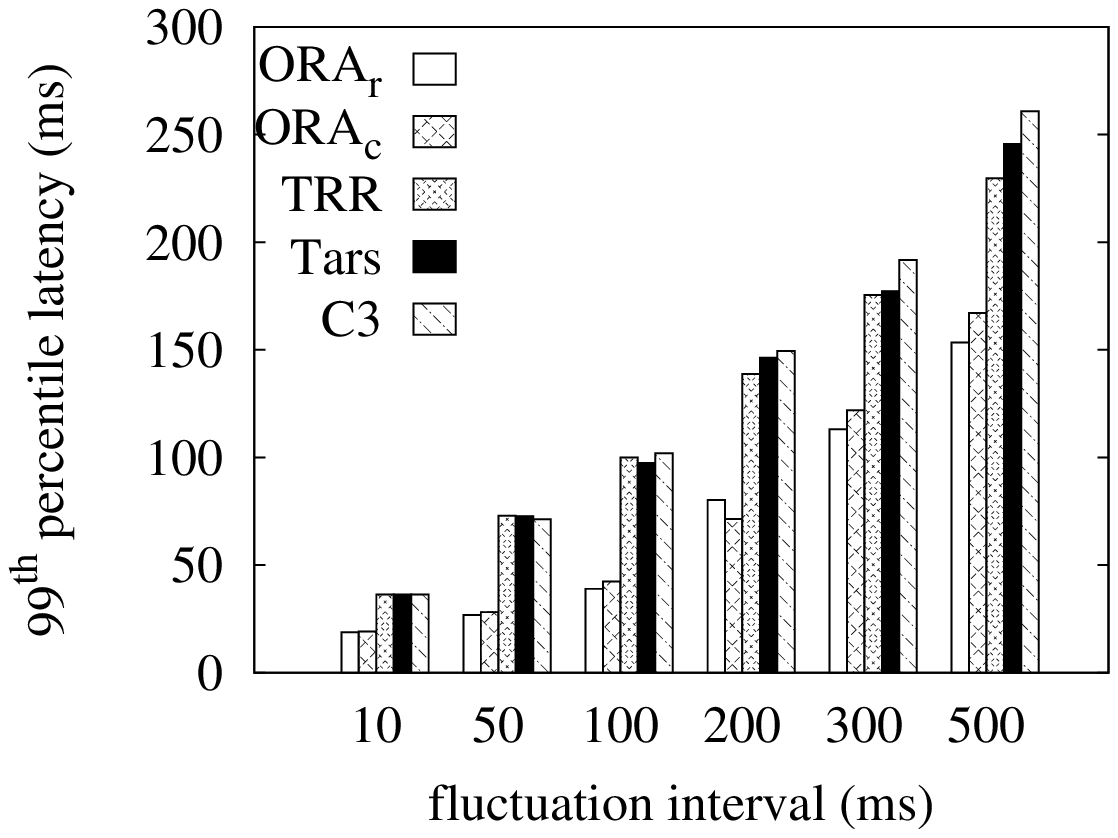}
}
\quad\quad\quad
\subfigure[n=300]{
\centering
\includegraphics[width=2.7in]{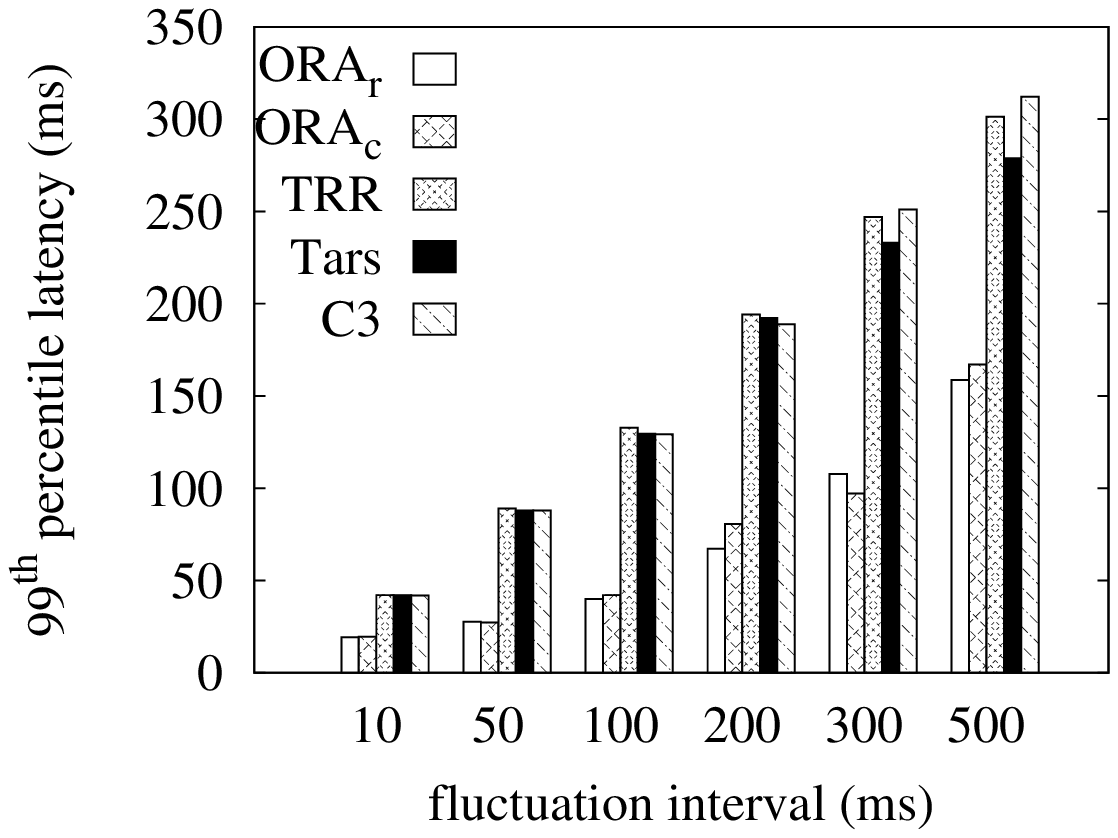}
}
\caption{Impacts of the skewed client demands: 50\% clients generate 80\% of the total demand.} \label{skew50}
\end{figure*}

\noindent\textbf{Impacts of the Number of Clients}
Subsequently, we increase the number of clients to be $n=300$ under the default high utilization scenario. The corresponding $99^{th}$ percentile latencies are shown in \figurename\ref{c300h}.
As discussed in the part C of section IV, the $\tau_w$ would has smaller probability to be of small values in this condition. This conclusion is confirmed by \figurename\ref{cdf300}, where the cumulative distribution function of $\tau_w$ with $n=300$ are presented, similar to \figurename\ref{waittime}.
When the $\tau_w$ is often of large values, the queue-size estimation will become worse and the rate adjustment result has to wait for a longer time before it takes effect.
Therefore, the $99^{th}$ percentile latencies illustrated in \figurename\ref{c300h} become larger than that in \figurename\ref{c150h}, respectively. But they have the same variation tendency with the change of the time interval  $T$.
Moreover, in these conditions, Tars also outperforms C3.

\noindent\textbf{Impacts of the Sever Utilization}
Next, we repeat above simulations under the low utilization scenario, where the arrival rate matches a 45\% server utilization.
The $99^{th}$ percentile latencies are shown in \figurename\ref{utilization}.
Comparing with above simulation results, the $99^{th}$ percentile latencies of both Tars and C3 are seldom influenced by the changes of the period, where the average service time changes, under the low utilization scenario.
Because once a server becomes slow according to the time-varying performance model, it is unlikely to be chosen by Tars and C3, as the other fast servers are unlikely saturated in this situation. 
Consequently, this slow server contributes little to the $99^{th}$ percentile latencies.
On the other hand, similar to above result, we can find the $99^{th}$ percentile latencies increase with the increase of the number of clients in \figurename\ref{utilization}.
In addition, Tars outperforms C3 in \figurename\ref{utilization}, especially when the number of clients becomes $n=300$.

\noindent\textbf{Impacts of the Skewed Demands}
As many realistic workloads are skewed in practice \cite{skew}, we evaluate Tars under the skewed client demands. Specifically, we respectively let 20\% or 50\% of the clients generate 80\% of the total keys towards the servers.
The $99^{th}$ percentile latencies are shown in \figurename\ref{skew20} and \figurename\ref{skew50}, respectively.
Consisting with above simulation results, Tars outperforms C3 under all of these two skewed demands scenarios.

In summary, Tars outperforms C3 under all kinds of conditions. The advantages of Tars over C3 is not very significant, because Tars is designed based on C3 with only a few modifications, and Tars is also unable to totally address the timeliness issue of the framework developed in C3.

\section{Conclusion and Further Work}
Nowadays, it is crucial to select the fastest replica server via the replica selection scheme, such that the tail latency of key-value accesses is reduced.
To address the challenges of the time-varying performance across servers and the herd behavior, an adaptive replica selection scheme C3 is proposed recently.
Despite of the innovations on bringing in the feedback for replica ranking and developing the rate control and backpressure mechanism, and the good performance of C3, we find drawbacks of C3 in respect of poor queue-size estimation and unsuitable goal of rate control, and reveal the timeliness issue of the framework developed by C3.
These insights motivate us to further develop the Tars scheme, improving the replica ranking by taking the timeliness of feedback information into account, and revising the goal of rate control.
Evaluation results based on the open source code of C3 confirm the good performance of Tars against C3.
Further work can be, but not limited to, evaluation of Tars with real experiments, totally addressing the timeliness issue of the framework developed by C3, and improvement of the rate control algorithm for key-value stores.


\begin{thebibliography}{1}
\bibitem{Dynamo}
G. Decandia, D. Hastorun, M. Jampani, G. Kakulapati, A. Lakshman, A. Pilchin, S. Sivasubramanian, P. Vosshall, and W. Vogels, \emph{Dynamo: Amazon’s Highly Available Key-value Store}, In Proc. of the SOSP, 2007.

\bibitem{bing}
V. Jalaparti, P. Bodik, S. Kandula, I. Menache, M. Rybalkin, and C. Yan, \emph{Speeding up Distributed Request-Response Workflows}, In Proc. of the SIGCOMM, 2013.

\bibitem{facebook}
R. Nishtala, H. Fugal, S. Grimm, M. Kwiatkowski, H. Lee, H. C. Li, R. McElroy, M. Paleczny, D. Peek, P. Saab, D. Stafford, T. Tung, and V. Venkataramani,\emph{Scaling Memcache at Facebook}, In Proc. of the NSDI, 2013.

\bibitem{latency}
S. M. Rumble, D. Ongaro, R. Stutsman, M. Rosenblum, and J. K. Ousterhout, \emph{Its time for low latency}, In Proc. of the HotOS, 2011.

\bibitem{AtScale}
J. Dean and L. A. Barroso, \emph{The Tail At Scale}, Communications of the ACM, Volumn 56:74-80, 2013.

\bibitem{revenue}
J. Brutlag, \emph{Speed Matters}, \url{http://googleresearch.blogspot.com/2009/06/speed-matters.html}, 2009

\bibitem{redundancy}
A. Vulimiri, P. B. Godfrey, R. Mittal, J. Sherry, S. Ratnasamy, and S. Shenker, \emph{Low Latency via Redundancy}, In Proc. of the CoNEXT, 2013.

\bibitem{CosTLO}
Z. Wu, C. Yu, and H. V. Madhyastha, \emph{CosTLO: Cost-Effective Redundancy for Lower Latency Variance on Cloud Storage Services}, In Proc. of the NSDI, 2015.

\bibitem{HDFS}
D. Borthakur, \emph{The hadoop distributed file system: Architecture and design}, Hadoop Project Website, 11(11):1-10, 2007.

\bibitem{Riak}
\emph{Riak Load Balancing and Proxy Configuration}, \url{http://docs.basho.com/riak/1.4.0/cookbooks/Load-Balancing-and-Proxy-Configuration/}, 2014.

\bibitem{Cassandra}
\emph{Cassandra Documentation}, \url{http://www.datastax.com/documentation/cassandra/2.0}, 2014.

\bibitem{C3}
L. Suresh, M. Canini, S. Schmid, and A. Feldmann, \emph{C3: Cutting Tail Latency in Cloud Data Stores
via Adaptive Replica Selection}, In Proc. of the NSDI, 2015.

\bibitem{Cubic}
S. Ha, I. Rhee, and L. Xu, \emph{CUBIC: A New TCP-Friendly High-Speed TCP Variant}, SIGOPS Oper. Syst. Rev., 42(5), 2008.

\bibitem{tcp}
V. Jacobson, \emph{Congestion Avoidance and Control}, In Proc. of the SIGCOMM, 1988

\bibitem{server-model}
J. Schad, J. Dittrich, and J.-A. Quian$\acute{e}$-Ruiz, \emph{Runtime Measurements in the Cloud: Observing, Analyzing, and Reducing Variance}, VLDB Endowment, 3(1-2), 2010


\bibitem{time-varying}
M. Kambadur, T. Moseley, R. Hank, and M. A. Kim, \emph{Measuring Interference Between Live Datacenter Applications}, In Proc. of the SC, 2012.

\bibitem{congestion}
K. Ousterhout, R. Rasti, S. Ratnasamy, S. Shenker and B. Chun, \emph{Making Sense of Performance in Data Analytics Frameworks}, In Proc. of the NSDI, 2015

\bibitem{mongodb}
K. Bogdanov, M. Peon-Quir$\acute{o}$s, G. Q. Maguire Jr. and D. Kosti$\acute{c}$, \emph{The Nearest Replica Can Be Farther Than You Think}, In Proc. of the SoCC, 2015

\bibitem{skew}
B. Atikoglu, Y. Xu, E. Frachtenberg, S. Jiang, and M. Paleczny, \emph{Workload Analysis of a Large-scale Key-value Store}, In Proc. of the SIGMETRICS, 2012

\end{thebibliography}
\end{document}